\documentstyle[aps,prl,twocolumn,epsf]{revtex}
\tighten
\begin{document}
\preprint{ITFA 95-24}
\draft
\title{Exact Solution of an Octagonal Random Tiling Model }
\author{Jan de Gier and Bernard Nienhuis}
\address{Instituut voor Theoretische Fysica, Universiteit van Amsterdam,
Valckenierstraat 65, \\ 1018 XE Amsterdam, The Netherlands. E-mail:
{\tt degier@phys.uva.nl}, {\tt nienhuis@phys.uva.nl}}
\address{\em \today}
\address{
\centering{
\medskip \em
\begin{minipage}{15.4cm}
{}~~~We consider the two-dimensional random tiling model introduced by Cockayne, i.e. the ensemble of all possible coverings of the plane without gaps or
overlaps with squares and various hexagons. At the appropriate
relative densities the correlations have eight-fold rotational
symmetry. We reformulate the model in terms of a random tiling
ensemble with identical rectangles and isosceles triangles. The
partition function of this model can be calculated by diagonalizing a
transfer matrix using the Bethe Ansatz (BA). The BA equations can be
solved providing {\em exact} values of the entropy and elastic constants. 
\end{minipage}
}}
\maketitle
\narrowtext

Since the discovery of quasicrystals, materials with
non-crystallographic rotational symmetry and quasiperiodic
translational order have been modelled by tilings. A tiling model
consists of a set of elementary building blocks, tiles, that cover
space without gaps or overlaps. One of the main questions concerning
quasicrystalline alloys 
is their thermodynamic stability. It has been argued by several
authors~\cite{Elser,Henley} that this may result partly from entropy
associated with local 
random rearrangements of the tiles. One is then naturally led to study
ensembles of 'random tilings'~\cite{Henley}. 

It is now known for some time~\cite{Widom,Kalugin} that in two
dimensions the square-triangle random tiling (RT) model, which has a
twelve-fold rotational symmetry, can be solved, i.e. its
entropy and phason elastic constants can be calculated
exactly. In this letter we give the results of such a calculation for
an eight-fold symmetric RT-model. 

The model under consideration consists of squares and hexagons of
arbitrary size, and was
first introduced by Cockayne \cite{Cockayne}. The hexagons are built
out of rectangles with sides $1:\sqrt{2}$ and a pair of isosceles and
rectangular triangles. The squares can be viewed as two triangles. The
model is therefore equivalent to a triangle-rectangle random tiling
with an extra Boltzmann-weight such that the two ways two triangles form a
square are counted as one,
i.e.:\\
\begin{picture}(160,20)(-30,-5)
\thicklines
\put(20,10){\line(1,1){10}}
\put(30,20){\line(1,-1){10}}
\put(20,10){\line(1,-1){10}}
\put(30,0){\line(1,1){10}}
\put(20,10){\line(1,0){20}}
\put(70,10){\line(1,1){10}}
\put(80,20){\line(1,-1){10}}
\put(70,10){\line(1,-1){10}}
\put(80,0){\line(1,1){10}}
\put(80,0){\line(0,1){20}}
\put(120,10){\line(1,1){10}}
\put(130,20){\line(1,-1){10}}
\put(120,10){\line(1,-1){10}}
\put(130,0){\line(1,1){10}}
\put(50,10){\line(1,0){10}}
\put(55,5){\line(0,1){10}}
\put(100,12){\line(1,0){10}}
\put(100,8){\line(1,0){10}}
\end{picture}\\
There is a repulsive potential of $k T \log 2$ for each pair of triangles 
adjacent by their long edge. Since the configuration of 
vertices does not depend on which diagonal is drawn in a square, the
partition sum is precisely the sum of all vertex configurations, 
rather than of all tilings.

It is interesting to note that the perfect quasicrystalline
square-hexagon tiling generated by an inflation rule \cite{Cockayne}
is in one-to-one correspondence with the binary octagonal tiling of squares and
rhombi. Although the random tiling ensemble of the latter set of tiles
has been studied, \cite{Li}, no exact solution in
the quasicrystalline phase has been found yet. 

\begin{figure}[h]
\begin{picture}(236,100)(0,0)
\put (-20,0){\epsffile{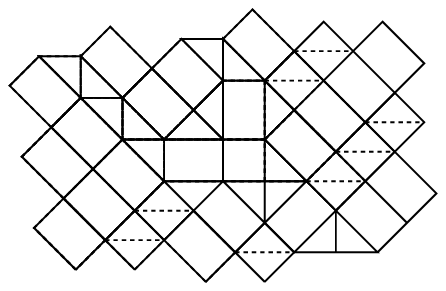}}
\put (110,0){\epsffile{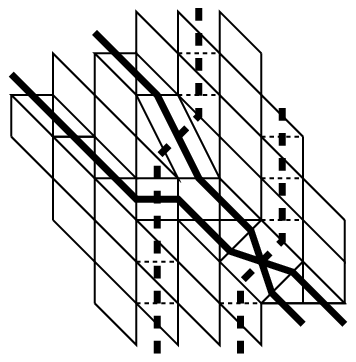}}
\end{picture}
\caption{(a) Patch of the tiling model. (b) Corresponding patch on the
  lattice. Bold solid lines are domain walls of horizontal short
  edges, referred to as $s$-walls in the text. Bold dashed lines
  represent the $l$-walls of horizontal long edges.}
\label{layerdecompo}
\end{figure}

Like in the square triangle tiling we can set up a transfer
matrix. This is done by decomposing the tiling into layers. Different
layers are bounded by the short horizontal edges, the horizontal
diagonals of the squares and the almost horizontal diagonals of the $\pm
{\pi\over 4}$ tilted rectangles. In addition, the layer edges cut the
triangles and rectangles with a vertical long edge in half. In this
way the tiles are deformed in such a way that the vertices of the
tiling form a subset of those of the
square lattice, see Fig.~\ref{layerdecompo}(b). The
horizontal diagonals of the squares are denoted by the dashed lines in
Fig.~\ref{layerdecompo}(a). A matrix element $T_{ij}$ of the transfer
matrix {\bf T} is $0$ if layer $j$ can not be followed by layer
$i$. Otherwise it is given by the statistical weight of the layer $i$.

In the following we will call the tilted
rectangles $R_{\pm}$ and the rectangles with the short and long
horizontal edge by $R_s$ and $R_l$ respectively. 

Because different tiles of the original tiling are mapped onto the same
shapes on the square lattice, we have to decorate the new configurations. This
is done with bold dashed and solid lines, see
Fig.~\ref{layerdecompo}(b). Thus it is clear that the horizontal short
and long edges of the original tiling form domain walls, which we denote by
type $s$ and $l$ respectively. Between two layer edges on the square
lattice, the $s$-walls step one unit to the left and the $l$-walls do
not move. When two walls cross, the $s$-wall may either jump over the
$l$-wall moving two places to the left and thereby creating a
rectangle $R_l$, or, over {\em two} layers, the walls may
exchange place creating a rectangle $R_s$. In the latter case the
  crossing therefore is completed after application of the
  transfermatrix twice. It may also happen that two
walls of type $s$ and 
one of type $l$ cross simultaneously over two layers. The
$l$-wall and the $s$-wall nearest to it then exchange place, while the
the second $s$-wall jumps over both these walls moving three places to
the left creating a rectangle $R_-$. 

We can express the tile densities in terms of the domain wall
densities. We shall denote the horizontal size of the tiling by $L$ and the
corresponding system size of the lattice model by $N$. Let
$\Delta_{ls}=R_l-R_s=0$, i.e. both types of collisions of two domain
walls occur
with the same frequency, and let $n_s$ and $n_l$ be the number of $s$-
and $l$-walls. Apply the
transfer matrix $p=2N-n_s$ times on some initial configuration of
domain walls at $t=0$ on the lattice, and suppose that both types of collisions
occur for every {\em pair} of $s$- and $l$-walls. The final state at
$t=2N-n_s$ will then be the same configuration of domain walls as the
initial one shifted by $n_l$. The total number of
rectangles and of triangles per layer can then be calculated to be:
\begin{eqnarray}
n_{rect}&\;=\;&N-n_s-n_l+2n_s n_l/p.\nonumber\\
n_{tri}&\;=\;&2(n_s+n_l)-6n_s n_l/p.
\end{eqnarray}
The tile densities that belong to the quasicrystalline phase are
$n_{rect}/N=6-4\sqrt{2},\;n_{tri}/N=12\sqrt{2}-16$, corresponding to an
area fraction of triangles $\alpha_t=1/2$. 

As a function of the domain wall densities, the model displays
two incommensurate phases. A 4-fold symmetric phase is formed in the high
density region, $\alpha_t>1/2$, where the triangles form octagonal and square
cells bounded by domain walls consisting of rectangles. There is a 2-fold
symmetric phase in the low density region where the rectangles from
rectangular cells bounded by the domain walls consisting of triangles.
  
Like the square-triangle tiling this tiling has the irrotational
property~\cite{Oxborrow} which implies that, according to the random
tiling hypotheses, the entropy density $\sigma_a=S/A$ has
the following form:
\begin{equation}
\sigma_a\;=\;\sigma_{a,0}-{1 \over 2}K_{\mu}\left({\rm Tr}{\bf E}\right)^2+{1
  \over 2}K_{\xi}\det{\bf E}+{\cal{O}}\left({\bf E}^3\right)\label{rthyp}
\end{equation}
where ${\bf E}$ is the phason strain tensor. The conditions on
  the elastic constants for ${\bf E}=0$ to be a local maximum are:
\begin{equation}
K_{\mu}>0,\;\;K_{\xi}>0,\;\;4K_{\mu}-K_{\xi}>0.\label{stable}
\end{equation}
We denote the deviations of
the ideal tile densities by
\begin{eqnarray} 
\delta_{ls}&\;=\;&n_l\sqrt{2}-n_s.\label{deltals}\\
\Delta_{\pm}&\;=\;&N-n_l-n_s.\label{Deltapm}
\end{eqnarray}
The quadratic forms in (\ref{rthyp}) can be expressed in these:
\begin{eqnarray}
\left({\rm Tr}{\bf E}\right)^2&\;=\;&{1\over L^2}\left(2\delta_{ls}-\Delta_{\pm}\left(2-\sqrt{2}\right)-\Delta_{ls}\left(1+\sqrt{2}\right)\right)^2.\nonumber\\ 
\det{\bf
  E}&\;=\;&{1\over L^2}\left(\delta_{ls}^2-\left(2-
\sqrt{2}\right)\Delta_{\pm}\delta_{ls}-2\Delta_{\pm}^2\sqrt{2}\right.\nonumber\\
&&\left.-\left(1+\sqrt{2}\right)\Delta_{ls}\left(\delta_{ls}-\left(1-1/\sqrt{2}\right)\Delta_{\pm}\right)\right).\nonumber\\\label{invariants}
\end{eqnarray}
The quantities $n_s$ and $n_l$ are conserved by the action of the
transfer matrix {\bf T}. To control the average value of
$\Delta_{ls}$, the tiles $R_s$
and $R_l$ are given a weight $\exp(-\phi)$ and $\exp(\phi)$
respectively. Furthermore, as the tiles $R_s$ and $R_+$ in the lattice
representation have an area that is twice that of the other two
transformed rectangles, we have to introduce a chemical potential for
them to compensate for this asymmetry. The tiles $R_s$ and $R_+$
therefore get an extra weight $\exp(\eta)$.  

The free energy per layer of the lattice model is given by the logarithm of the
largest eigenvalue of ${\bf T}$:
\begin{eqnarray}
F\left(n_s,n_l,\phi\right)&\;=\;&-\log\Lambda\nonumber\\
&\;=\;&-S-\phi\Delta_{ls}\nonumber\\
&&-\eta\left(n_{R_s}+n_{R_+}\right).
\end{eqnarray}

We denote the horizontal coordinate of the $i$th $s$-wall by $\xi_i$ and
of the $k$th $l$-wall by $z_k$. The vertical coordinate is denoted by
by $t$. Let $l_i$ be the total number of
$l$-walls to the left of the $i$th $s$-wall, then the quantity
$\xi_i+t+l_i \bmod 2$ is conserved for every $s$-wall. This means that
the $s$-walls lie on a sublattice structure and split up into two
kinds: odd and even ones. Denote their coordinates by $x_i$ and $y_j$
respectively.

The eigenvectors of ${\bf T}$ as a function of the coordinates $x$,
$y$ and $z$ of the domain walls are of the Bethe-Ansatz form. If
  the domain walls all are separated the Ansatz is:
\begin{equation}
\sum_{\pi, \mu, \rho} A\left(\Gamma\right)\prod_{i=1}^{n_{s,o}}
u_{\pi_i}^{x_i}\prod_{j=1}^{n_{s,e}}
v_{\mu_j}^{y_j}\prod_{k=1}^{n_l} w_{\rho_k}^{z_k}.\label{eigvec}
\end{equation}
 The form of the eigenvector for configurations where domain
  walls cross can be found by application of
{\bf T} on (\ref{eigvec}).
Here, $w_k=\exp({\rm i} q_k)$, $u_i=\exp({\rm i} p_{o,i})$ and
$v_j=\exp({\rm i} p_{e,j})$ are the exponentiated momenta and
$\rho$, $\pi$ and $\mu$ are the permutations of these belonging to
the $l$- and odd and even $s$-walls respectively. The amplitudes $A$
depend on the permutations $\rho$, $\pi$ and $\mu$ and on the
configuration of the various domain walls. These together are
coded in a vector $\Gamma$ in the following way. Let
{\bf r} be the vector of coordinates $x_i$, $y_j$ and $z_k$ of all domain
walls, ordered so
that $r_m < r_{m+1}$. The entries of $\Gamma$ are the elements of the
permutations $\pi$, $\mu$ and $\rho$. The
order of succesion in $\Gamma$ of elements taken from $\pi$, $\mu$ and
$\rho$ matches that of the elements of $x$, $y$ and $z$ respectively in
{\bf r}. So, for example, in the case of an odd
$s$-wall at $x_1$ and an $l$-wall at $z_1$ we would either have $x_1 <
z_1$, or $x_1 > z_1$. In the first case we write ${\bf
  r}=\left(x_1,z_1\right)$ with $\Gamma = \left(\pi_1,\rho_1\right)$,
while in the second case ${\bf r}=\left(z_1,x_1\right)$ and $\Gamma
=\left(\rho_1,\pi_1\right)$. 

When the different domain walls
are separated, the transfer matrix shifts all $s$-walls to
the left and leaves all $l$-walls at rest, so the eigenvalue of ${\bf
  T}$ must be  
\begin{equation}
\Lambda\;=\;\prod_{i=1}^{n_{s,o}}u_i\prod_{j=1}^{n_{s,e}}v_j.
\end{equation}
Inspecting the eigenvalue equations for the case that an $s$- and an
$l$-domain wall collide one sees that the amplitudes $A$ before and
after the collision must satisfy
the following relation for (\ref{eigvec}) to be an eigenvector of
${\bf T}$:
\begin{equation}
{A\left(\dots\pi_i,\rho_k\dots\right) \over
  A\left(\dots\rho_k,\pi_i\dots\right)}\;=\;\left({\rm
  e}^{\phi}u_{\pi_i}+{\rm e}^{\eta-\phi}u_{\pi_i}^{-1}w_{\rho_k}^{-1}\right).
\end{equation}
A same relation holds for the amplitudes with $\pi_i$ replaced by
$\mu_j$ and $u$ replaced by $v$. From configurations involving
three domain walls one deduces that interchanging domain walls of
the same kind in the amplitude gives a factor $-1$ and that
interchanging an odd and an even $s$-wall leaves the
amplitude unchanged. The eigenvalue equations therefore do not mix the
momenta of the even and odd $s$-walls. It turns out that all relations
among amplitudes involving more than two domain walls factorize into
the ones already mentioned. These relations therefore suffice to make
(\ref{eigvec}) an eigenvector of {\bf T}. 
Imposing periodic boundary conditions and eliminating the amplitudes
$A$ from the eigenvalue equations one gets the following equations for
the momenta:
\begin{eqnarray}
u_i^L&=&(-1)^{n_{s,o}-1}\prod_{k=1}^{n_l}\left({\rm e}^{\phi}u_i+{\rm
  e}^{\eta-\phi}u_i^{-1}w_k^{-1}\right),\nonumber\\
v_j^L&=&(-1)^{n_{s,e}-1}\prod_{k=1}^{n_l}\left({\rm e}^{\phi}v_j+{\rm
  e}^{\eta-\phi}v_j^{-1}w_k^{-1}\right),\nonumber\\
w_k^{-L}&=&(-1)^{n_l-1}\prod_{i=1}^{n_{s,o}}\left({\rm
  e}^{\phi}u_i+{\rm
  e}^{\eta-\phi}u_i^{-1}w_k^{-1}\right)\times\nonumber\\
&&\times\prod_{j=1}^{n_{s,e}}\left({\rm e}^{\phi}v_j+{\rm
  e}^{\eta-\phi}v_j^{-1}w_k^{-1}\right).\label{BAE}
\end{eqnarray}
These are the so-called Bethe Ansatz equations (BAE). Like the BAE for
the square-triangle tiling these equations can be solved along a line in the
thermodynamic limit for the largest eigenvalue. The details of this
calculation, which resembles the one by Kalugin for the
square-triangle tiling \cite{Kalugin}, will be published
elsewhere. Here we only give the results.

\begin{figure}[h]
\begin{picture}(236,160)(0,0)
\put (0,0){\epsffile{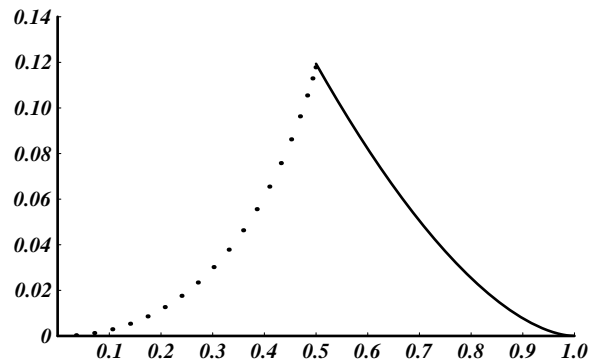}}
\end{picture}
\caption{$\sigma_a$ as a function of $\alpha_t$. The solid line
  corresponds to the exact solution (\ref{exsol}). The dots are
  numerical results for $N=198$.}
\label{sa}
\end{figure}

The entropy can be calculated exactly in the entire regime
$\alpha_t\geq 1/2,\;\Delta_{\pm}=\Delta_{ls}=0$. In this region the
three curves formed by the
solutions of the BAE have the same limitpoint. After a change of
variables, this limitpoint can be written in the notation of Kalugin
as $b={\rm i} |b|{\rm e} ^{{\rm i}\gamma}$. The tile densities and the
area fraction can be expressed in $\gamma$.
\begin{eqnarray}
n_l/N&\;=\;&1-n_s/N\;={1-\sqrt{2}\sin\gamma/2\over
  1+\sqrt{2}\cos\gamma/2}.\label{density}\\
\alpha_t&\;=\;&\left(\sqrt{2}+1\right){\sqrt{2}-
  \cos\gamma/2\over1+\cos\gamma/2}.
\end{eqnarray}
The entropy per area of the square-hexagon random tiling in the regime
$1/2 \leq \alpha_t \leq 1$ in terms of $\gamma$ is:
\begin{eqnarray}
\sigma_a&\;=\;&{\sqrt{2}+1\over
  4\left(\cos{\gamma/2}+1\right)}\left(2\sqrt{2}\log{4\over
  \cos\gamma}\right.\nonumber\\
&&\left.-\left(\cos{\gamma/2}-\sin{\gamma/2}\right)\log\left({1+
  \cos\left({\pi/4}+{\gamma/2}\right)\over
  1-\cos\left({\pi/4}+{\gamma/2}\right)}\right)\right.\nonumber\\
&&\left.
-\left(\cos{\gamma/2}+\sin{\gamma/2}\right)\log\left({1+
  \cos\left({\pi/4}-{\gamma/2}\right)\over
  1-\cos\left({\pi/4}-{\gamma/2}\right)}\right)\right).\nonumber\\\label{exsol}
\end{eqnarray}
The entropy has its maximum at $\gamma=0$. Expanding $\sigma_a$ up to
second order in $\gamma$ results in:
\begin{equation} 
\sigma_a\;=\;\sigma_{a,0}-\gamma^2{1+\sqrt{2}\over
  32\sqrt{2}}\left(4-\log
4-\sqrt{2}\log\left(1+\sqrt{2}\right)\right),\label{gamexp}
\end{equation}
where the residual entropy at $\gamma=0$ is given by
\begin{eqnarray}
\sigma_{a,0}&\;=\;&{1+\sqrt{2}\over2\sqrt{2}}\left(\log
4-\sqrt{2}\log\left(1+\sqrt{2}\right)\right)\nonumber\\
&\;\approx\;&0.1193642186\dots.
\end{eqnarray}
As in the square triangle tiling, the entropy is a convex function of
the area fraction $\alpha_t$, see Fig.~\ref{sa}.

In the 2-fold phase $(\alpha_t < 1/2)$ this exact calculation fails
because the solution curves do not have the same
limitpoint. Nevertheless, it is
possible to calculate the lowest order correction to the entropy. With
$\epsilon=\Delta_{\pm}/N$ this is given by
\begin{eqnarray}
\sigma_a &\;=\;&\sigma_{a,0}-\epsilon\gamma{1\over
  16\sqrt{2}}\left(4-\log{4}-\sqrt{2}\log\left(1+\sqrt{2}\right)\right)\nonumber\\
&&-\epsilon^2{1+\sqrt{2}\over 8}\left(\log
4+\sqrt{2}\log\left(1+\sqrt{2}\right)\right).\label{epsexp}
\end{eqnarray}
This expression gives the exact slope of the numerical curve
  shown in Fig. \ref{sa} for $\alpha_t\uparrow 1/2$:
\begin{equation}
\left.{d\sigma_a\over
    d\alpha_t}\right|_{\alpha_t \uparrow {1\over 2}}\;=\;{\sqrt{2}-1\over
    \sqrt{2}}\left(\log{4}+\sqrt{2}\log\left(1+\sqrt{2}\right)\right).
\end{equation}
Expanding equation (\ref{density}) up to first order in $\gamma$ and using
(\ref{rthyp})$-$(\ref{invariants}), it is straightforward to find the
elastic constants $K_{\mu}$ and $K_{\xi}$ from (\ref{gamexp}) and (\ref{epsexp}). Their numerical values are:
\begin{equation}
K_{\mu}\;=\;0.2842712\dots,\;\;K_{\xi}\;=\;0.7366252\dots.
\end{equation}
Since the two elastic constants fulfill the relations (\ref{stable})
the quasiperiodic eightfold symmetric state is entropically stable. 

In this paper we succesfully apply the Bethe Ansatz to an
octagonal random tiling model. The BAE~(\ref{BAE}) are solved to find
{\em exact} values of the entropy and elastic constants. The model
shows qualitatively the same behaviour as the square-triangle
tiling. It is not yet clear how generic the solvability of these two
tilings is, but we have discovered that a 10-fold symmetric tiling of
rectangles and triangles does admit a Bethe Ansatz. It appears however
from numerical calculations that their solutions do not allow for an
exact solution using the method of Kalugin which is employed in this paper.

We like to thank Chris Henley and Mike Widom for providing us with
some useful information. This work was supported by FOM, part of NWO, Institute
for Dutch Scientific Research.


\begin{references}
\bibitem{Elser} V. Elser, Phys. Rev. Lett. {\bf 54}, 1730 (1985).
\bibitem{Henley} See C.L. Henley in {\em Quasicrystals: The State of
    the Art}, edited by P.J. Steinhardt and D.P. DiVincenzo (World
  Scientific, Singapore, 1991) p 429 for a review of random tiling concepts.
\bibitem{Widom} M. Widom, Phys. Rev. Lett. {\bf 70}, 2094 (1993).
\bibitem{Kalugin} P.A. Kalugin, J. Phys. A:Math Gen. {\bf 27} 3599
  (1994).
\bibitem{Cockayne} E. Cockayne, J. Phys. A:Math. Gen. {\bf 27} 6107
  (1994).
\bibitem{Li} W. Li, H. Park and M. Widom, J. Stat. Phys. {\bf 66}, 1 (1992).
\bibitem{Oxborrow} M. Oxborrow and C.L. Henley, Phys. Rev. B {\bf 48}
  6966 (1993).
\end{references}
\end{document}